\documentclass[pra,twocolumn,showpacs,superscriptaddress]{revtex4-1}
\usepackage[colorlinks, linkcolor=blue, citecolor=blue, urlcolor=blue, breaklinks=true]{hyperref}

\usepackage{multirow,eurosym,amssymb,amsfonts,amsmath,setspace,graphicx,color,bm,float,verbatim}

\newcommand{\bra}[1]{\langle #1|}
\newcommand{\ket}[1]{|#1\rangle}

\def\ket#1{| #1 \rangle}
\def\bra#1{\langle #1 |}

\def\be{\begin{equation}}
\def\ee{\end{equation}}

\def\bsplit{\begin{split}}
\def\nsplit{\end{split}}

\begin{document}
\title{Loss-resilient photonic entanglement swapping using optical hybrid states}
\date{\today}

\author{Youngrong Lim}
\affiliation{Center for Macroscopic Quantum Control, Department of Physics and Astronomy, Seoul National University, Seoul 151-742, Republic of Korea}

\author{Jaewoo Joo}
\affiliation{School of Computational Sciences, Korea Institute for Advanced Study, Seoul 02455, Republic of Korea}

\author{Timothy P. Spiller}
\affiliation{York Centre for Quantum Technologies, Department of Physics, University of York, York YO10 5DD, U.K.}

\author{Hyunseok Jeong}
\affiliation{Center for Macroscopic Quantum Control, Department of Physics and Astronomy, Seoul National University, Seoul 151-742, Republic of Korea}

\begin{abstract}
We propose a scheme of loss-resilient entanglement swapping between two distant parties via an imperfect optical channel. In this scheme, two copies of hybrid entangled states are prepared and the continuous-variable parts propagate through lossy media. 
In order to perform successful entanglement swapping, several  different measurement schemes are considered for the continuous-variable parts such as single-photon detection for ideal cases and a homodyne detection 
for practical cases. We find that the entanglement swapping using hybrid states with small amplitudes 
offers larger entanglement than the discrete-variable entanglement swapping in the presence of large losses. Remarkably, this hybrid scheme still offers excellent robustness of entanglement to the detection inefficiency. Thus, the proposed scheme could be used for the practical quantum key distribution in hybrid optical states under photon losses.
\end{abstract}


\maketitle
\section{Introduction}
\label{Sec1}
A distribution of entanglement at distance is one of the essential challenges for the practical schemes of quantum communication such as quantum key distribution~\cite{Ekert,QKD} and quantum secret sharing~\cite{secret}. The idea of entanglement swapping is particularly useful for long-distant quantum communication  (e.g., using quantum repeaters~\cite{repeater}). In a entanglement swapping scheme, successful joint measurements are used to guarantee faithful entanglement sharing between distant parties~\cite{Zukowski,esreview}. 

We start with the assumption that two communication parties are far separate from each other. Each of them prepares a bipartite entangled state independently, and sends one of the two qubits (a traveling qubit) to the middle location. After obtaining successful outcomes of a Bell-state measurement (BSM) in the middle, an entangled state is successfully shared by the two separate parties. In practice, quality of the shared entanglement degrades because of the imperfection of channels and detections. A noise-resilient entanglement swapping scheme is thus important for practical quantum communication in a realistic lossy channel. 

The optical implementation has provided the best platform for practical entanglement swapping and many experiments have been demonstrated in discrete-variables (DVs), e.g., photon polarization entangled states~\cite{DV1,DV2} and vacuum-single-photon (VSP) entangled states~\cite{DV3} as well as in continuous-variables (CVs), e.g., squeezed states~\cite{CV1,CV2} and coherent states~\cite{CV3, CV4, CV5}. In this work, our aim is to have the VSP entangled state, $\ket{\phi^+_{01}} = (\ket{0}\ket{1}+\ket{1}\ket{0})/\sqrt{2}$, where $\ket{0}$ is a vacuum state and $\ket{1}$ is a single photon state, shared by two distant parties. In fact, the VSP entangled state can be useful for quantum teleportation~\cite{Lee}, single-rail logic quantum computation~\cite{Ralph,Ralph05}, and single-photon nonlocality tests~\cite{Collett, Hardy,BH2004}.

A simple method of entanglement swapping for $\ket{\phi^+_{01}}$ is performed with two VSP entangled states and an all-optical BSM.
The traveling photons arrive at the middle location and are projected onto the VSP entangled states to build a BSM. This creates  a maximally entangled state between two distant parties, for which we shall present details in Section \ref{Sec2-B}. However, the success probability of a BSM is bounded by $1/2$ when using only linear optical elements~\cite{50}.
In contrast, a BSM based on CV qubits and entangled coherent states can achieve a larger success probability over this limit \cite{Jeong01,CV3}.
We shall thus attempt to use hybrid entanglement for entanglement swapping to obtain VSP entangled states shared by two distant parties.

In this paper, we are interested in utilizing hybrid entanglement in the form of
\begin{eqnarray} 
\ket{\psi_{HE}}_{AB} &=&  {1 \over \sqrt{2}} \left( \ket{0}_A \ket{ \alpha}_B + \ket{1}_A \ket{-\alpha}_B  \right), ~ 
\label{Psi_Ent01} 
\end{eqnarray} 
where $\ket{ \pm\alpha}$ are coherent states with amplitudes $\pm\alpha$.
This type of state can be generated using a cross-Kerr nonlinearity~\cite{hybrid1,hybrid2,hybrid3} but there are fundamental limitations to implement this interaction~\cite{prob1,prob2,prob3}.
Recently, this state was experimentally generated using linear optics elements without Kerr-type nonlinear interaction \cite{Jeong2014,Morin2014}. This kind of optical hybrid entanglement is useful to the quantum key distribution protocols~\cite{QKD1,QKD2}, quantum repeater~\cite{repeater1,repeater2},  quantum teleportation~\cite{Lee2013,SLB,HYK}, quantum computation with near-deterministic gate operations~\cite{Lee2013}, and Bell inequality test~\cite{Kwon2013}. In this hybrid entanglement swapping, the propagating parts are both CV qubits and the successful entanglement swapping eventually builds a DV entangled state. Our hybrid entanglement swapping scheme with the small $\alpha$ shows the advantages of robustness in the presence of photon losses.

This manuscript is organized as follows. In Sec.~\ref{Sec2}, we introduce a background with notations for the no-loss DV entanglement swapping scheme. Then, we describe how to perform the hybrid entanglement swapping in a lossy channel and investigate the ideal and practical entanglement swapping schemes in Sec.~\ref{Sec3} and \ref{Sec4-A}. In Sec.~\ref{Sec4-B}, we consider the inefficient detections in the presence of losses and provide the remarks in Sec.~\ref{Sec5}.

\section{Background}

\subsection{ Beam splitter operation}
\label{Sec2-A}

The beam splitter (BS) operation is a key element in our scheme for BSM in the middle location. It is also useful for the description of lossy channels. A general BS operator for two modes is given by
\begin{eqnarray}
\hat{B}^T (\theta,\phi)=\exp \left[i{\theta \over 2}(\hat{a}^{\dag}\hat{b}e^{i\phi}-\hat{a}\hat{b}^{\dag}e^{-i\phi}) \right]
\end{eqnarray}
with the transmission rate $T=\cos^{2} (\theta / 2)$~\cite{BS}, where $\hat{a}^{\dag}$ and $\hat{b}^{\dag}$ ($\hat{a}$ and $\hat{b}$) are the creation (annihilation) operators for each mode. If we set the phase $\phi=\pi$ and $\theta={\pi / 2}$ for a 50:50 BS ($BS^{1/2}$), the single photon or vacuum states after the BS are given by
\begin{eqnarray}
&& BS^{1/2}_{A,B} \, \ket{1}_A \ket{0}_B = (\ket{1}_A \ket{0}_B + \ket{0}_A \ket{1}_B) /\sqrt{2}, \label{BS01_1} \\
&& BS^{1/2}_{A,B} \, \ket{0}_A \ket{1}_B = (\ket{0}_A \ket{1}_B - \ket{1}_A \ket{0}_B) /\sqrt{2}, \label{BS01_2}\\
&& BS^{1/2}_{A,B} \, \ket{1}_A \ket{1}_B = (\ket{0}_A \ket{2}_B - \ket{2}_A \ket{0}_B) /\sqrt{2}, \label{BS01_3}
\end{eqnarray}
where $\ket{0}$, $\ket{1}$, and $\ket{2}$ are vacuum, single-photon, and two-photon Fock states respectively.

The state after $BS^{1/2}$ acting on the two coherent states is
\begin{eqnarray} 
BS^{1/2}_{A,B} \, \ket{\alpha}_A \ket{\beta}_B 
= \ket{{\alpha-\beta \over \sqrt{2} } }_A \,
\ket{{\alpha + \beta \over \sqrt{2} } }_B, \label{Back_02} 
\end{eqnarray}
and some cases of two input coherent states with the same value of the absolute amplitudes are  
\begin{eqnarray} 
&& BS^{1/2}_{A,B} \, \ket{\alpha}_A \ket{\alpha}_B = \ket{0}_A \ket{\sqrt{2} \alpha }_B, \label{SimpleBS_01} \\
&& BS^{1/2}_{A,B} \, \ket{ \alpha}_A \ket{- \alpha}_B  = \ket{ \sqrt{2} \alpha}_A \ket{0 }_B , \label{SimpleBS_02} \\
&& BS^{1/2}_{A,B} \, \ket{- \alpha}_A \ket{ \alpha}_B  = \ket{- \sqrt{2} \alpha}_A \ket{0 }_B , \label{SimpleBS_02-2} \\
&& BS^{1/2}_{A,B} \, \ket{-\alpha}_A \ket{-\alpha}_B = \ket{0}_A \ket{-\sqrt{2} \alpha}_B. \label{SimpleBS_03}
\end{eqnarray}

\subsection{Measurement schemes for CV BSMs}
\label{Sec2-C}
We shall  consider several detection schemes such as single-photon detection, photon on-off and homodyne measurements for CV BSMs. 
The ideal description of the photon number detection is a photon-number resolving (PNR) detector~\cite{PNR1,PNR2,PNR3} given by the $n$-photon projectors $\hat{M}_{\rm PNR} = \left\{ \sum_n \hat{P}^{j}_{n}= \ket{n}_{j}\bra{n} \right\}$. However, its implementation is in general very difficult and the PNR measurement consists in practice of many BSs and several photon on-off detectors given by
\begin{eqnarray} 
&& \hat{M}_{\rm on/off} = \left\{ \hat{P}^{j}_0, ~~\hat{P}^{j}_{\ne 0} = \openone - \ket{0}_{j}\bra{0} \right\},
\label{on-off_01}
\end{eqnarray} 
for spatial mode $j$ \cite{threshold_detect01}.
A single-photon detector is described by
\begin{eqnarray} 
&& \hat{M}_{\rm SPD} = \left\{ \hat{P}^{j}_0,~~\hat{P}^{j}_1 ,~~\hat{P}^{j}_{\ne 0,1} = \openone - \sum_{n=0,1}\ket{n}_{j}\bra{n} \right\}.~~~~~
\label{Thres_01}
\end{eqnarray} 

Alternatively, a reliable setup of homodyne detection is commonly used for CV photonic qubits and consists of a $BS^{1/2}$, a strong coherent field $\ket{\beta\,e^{i\theta}}$ and two photodetectors \cite{Peter05}. If the homodyne measurement is performed on a input signal in mode $B1$, the coherent field is injected with amplitude $\beta$ in mode $B2$ and the $BS^{1/2}_{B1,B2}$ mixes the input state and the field. The intensity difference between the two detectors located in the output fields is given by $I_{B1-B2}= \hat{b}_1 \hat{b}^{\dag}_2 - \hat{b}^{\dag}_1 \hat{b}_2 $ for creation operator $\hat{b}^{\dag}_{i}$ in $Bi$,
\begin{eqnarray} 
I_{B1-B2} &=&  2 |\beta| \langle \hat{x}_{\theta} \rangle, ~ 
\label{Homodyne_01} 
\end{eqnarray}
for $ \hat{x}_{\theta} = (\hat{b}_1 e^{i \theta} + \hat{b}^{\dag}_1 e^{-i \theta})/2$ \cite{Peter05,Andersen13,Suzuki06,Tim05}.
Note that a projection operator in mode $j$ is equal to
\begin{eqnarray} 
\hat{P}^{x_{\theta}}_{j} =  |  x_\theta\rangle_{j} \langle x_\theta |,
\label{Homodyne_04} 
\end{eqnarray}
for axis angle $\theta$ in the phase space and the probability amplitude is given by
\begin{eqnarray} 
\langle x_\theta | \alpha e^{i \varphi} \rangle 
&=& {1\over \pi^{{1\over 4}}} \exp \Big[ -{1\over 2} (x_{\theta})^2 + \sqrt{2} e^{i(\varphi-\theta)} \alpha x_{\theta}~~~~~ \nonumber \\
&& ~~~~~~~~~~~ - {1\over 2} e^{2i(\varphi-\theta)} \alpha^2 -  {1\over 2} \alpha^2 \Big]. ~ 
\label{Homodyne_02} 
\end{eqnarray}

\label{Sec2}
\begin{figure}[b]
\includegraphics[width=\columnwidth]{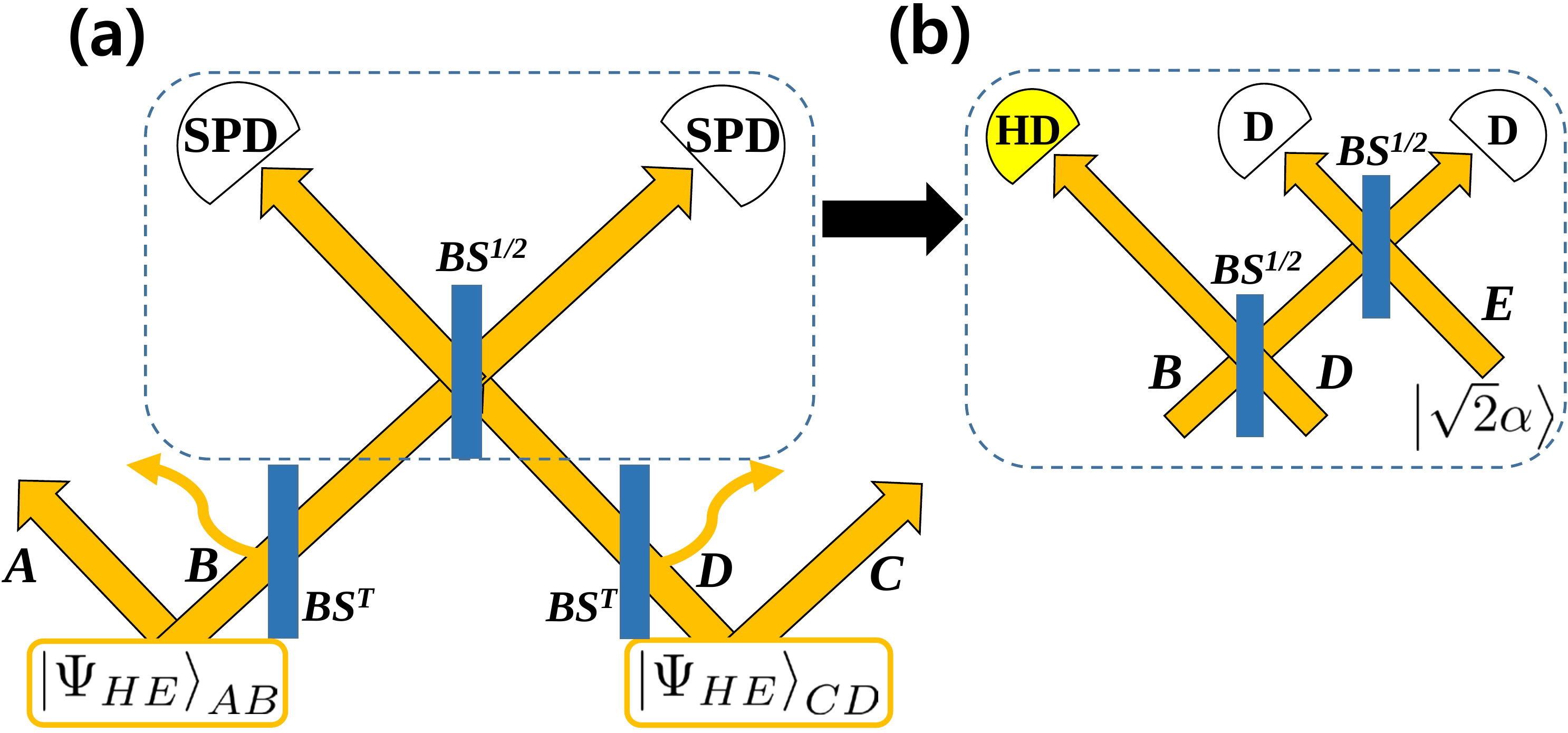}
\caption{(Color online) Schematic diagram of hybrid entanglement swapping schemes {\bf (a)} with single-photon detectors (SPD) and {\bf (b)} with a homodyne detector (HD) and on-off detectors (D). The hybrid states $\ket{\psi_{HE}}_{AB}$ and $\ket{\psi_{HE}}_{CD}$ with amplitudes $\alpha$ are prepared and the traveling coherent-state parts ($B$ and $D$) suffer   photon losses  described by $BS^{T}$.
In scheme (a), two single photon detectors are used together with  $BS^{1/2}$ for a BSM for coherent state qubits \cite{CV3, Jeong01}. In scheme (b), the part for the BSM is replaced with a homodyne detector after $BS^{1/2}$, an extra coherent state with amplitude $\sqrt{2}\alpha$, another 1:1 beam splitter and two on-off photodetectors as shown in panel (b).}
\label{scheme}
\end{figure}

\subsection{Entanglement swapping schemes without losses}
\label{Sec2-B}
Let us describe here the details of DV entanglement swapping using VSP entangled states. 
For the DV entanglement swapping, the initial state is prepared in  $\ket{\Psi_{DV}}_{ABCD}=\ket{\phi^{+}_{01}}_{AB}\ket{\phi^{+}_{01}}_{CD}$ with the traveling modes in $B$ and $D$. 
After the $BS^{1/2}_{BD}$, the DV BSM should be measured by projectors $ \{ \ket{\psi^{\pm}_{01}} \bra{\psi^{\pm}_{01}}, \ket{\phi^{\pm}_{01}} \bra{\phi^{\pm}_{01}} \}$ with
\begin{eqnarray} 
\label{VSP_01} 
\ket{\psi^{\pm}_{01}} =\left( \ket{0} \ket{0} \pm \ket{1} \ket{1}  \right)/\sqrt{2}, \\
\ket{\phi^{\pm}_{01}} =\left( \ket{0} \ket{1} \pm \ket{1} \ket{0}  \right)/\sqrt{2}.
\label{VSP_02} 
\end{eqnarray}
However, it is known that photonic DV BSM cannot be performed with a unit success probability using only linear optical elements in modes $B$ and $D$ (even without a photon loss in the channel). 

The success cases of the DV BSM are described by the two projection operators 
\begin{eqnarray}
&&\hat{P}_{01} =\hat{P}^{B}_{0} \otimes \hat{P}^{D}_{1} = \ket{0}_{B}\bra{0}\otimes\ket{1}_{D}\bra{1}, \label{proj_01_01}~~~~~~\\
&&\hat{P}_{10} =\hat{P}^{B}_{1} \otimes \hat{P}^{D}_{0}  =\ket{1}_{B}\bra{1}\otimes\ket{0}_{D}\bra{0},
\label{proj_01_02}
\end{eqnarray}
which can be performed by two single-photon detectors. Each projection operator $\hat{P}_{01} (\hat{P}_{10})$ brings the final state of DV entanglement swapping as $\ket{\phi^{+}_{01}}_{AC} (\ket{\phi^{-}_{01}}_{AC})$ respectively. Therefore, the outcome is also a maximally entangled state if the DV BSM has been successful performed.

On the other hand, a photonic CV entanglement swapping is quite different from the DV case because the CV BSM with coherent-state qubits can be performed in a nearly perfect manner if photon number parity measurements are possible \cite{CV3, Jeong01}. Namely, all four CV Bell states
\begin{eqnarray}
&&\ket{\Phi^{\pm}}=N_{\pm}(\ket{\alpha}\ket{\alpha}\pm \ket{-\alpha}\ket{-\alpha}),\nonumber\\
&&\ket{\Psi^{\pm}}=N_{\pm}(\ket{\alpha}\ket{-\alpha}\pm \ket{-\alpha}\ket{\alpha}),
\end{eqnarray}
where $N_{\pm}=(2\pm2e^{-4|\alpha|^2})^{-1/2}$, can be discriminated using a 50:50 beam splitter and two PNR detectors because only one of the two detectors registers photon(s) with a definite parity (even or odd) while the other registers no photon \cite{CV3,Jeong01}.
Even though there is a failure probability $\approx(2\cosh2\alpha^2)^{-1}$~\cite{Lee2015} for which both the detectors are silent, this probability rapidly approaches zero as $\alpha$ increases. 
Thus, without photon losses during the channel transmissions, the CV entanglement swapping generally provides higher success probabilities than the DV entanglement swapping does although both the cases can give maximum entanglement.
If PNR detectors for the CV BSM are unavailable, an efficient strategy particularly with small amplitudes is to use single photon detectors that discriminate between `zero', `one' and `two or more' photon(s) as those were used for the tele-amplification protocol~\cite{Sasaki2013}. In order to reflect more realistic conditions, we shall employ this method with single photon detectors for the CV BSM for our entanglement swapping scheme as depicted in Fig.~1(a).

When channel losses are present, the parity measurement for CV qubits and the single-photon measurement for DV qubits suffer the defects of distinguishability and the resultant states become mixed states with less entanglement.
To compare the degrees of entanglement of mixed states produced by different entanglement swapping schemes, we use an entanglement measure called negativity~\cite{nega} given by
\begin{eqnarray} 
E (\rho_{AC}) =-2\sum_{i} \lambda_{i}^{-},
\end{eqnarray}
where $\lambda_{i}^{-}$'s are 
negative eigenvalues of the partial transpose of $\rho_{AC}$. Here, $E=1$ is for a maximally entangled state while $E=0$ implies no entanglement at all.

\section{entanglement swapping schemes with a lossy channel}
\label{Sec3}
In this section, we consider a lossy environment  and compare the hybrid entanglement swapping with small $\alpha$ and the DV entanglement swapping, noting that single-photon detectors can be used in the BSM for both the DV and the hybrid entanglement swapping schemes. At the end of the section, we show that the homodyne detection, well-established for CV qubits in quantum optics, can be also used for the BSM of the hybrid case with assistance of a POVM as shown in Fig.~1(b).

\subsection{DV entanglement swapping in a lossy environment}
\label{Sec3-A}
For the comparison between the cases of hybrid and DV entanglement swappings, we first investigate the DV case in a lossy channel. The initial state is $\ket{\Psi_{DV}}_{ABCD}$ and the photons in mode $B$ and $D$ travel via photon-lossy channels modeled by $BS^{T}$ where $T$ is the transmission rate of the BS. In details, this approach is described by the reflection rate $R=1-T=1-e^{-\gamma\tau}$ where $\gamma$ is the decoherence rate and $\tau$ is the interaction time in the master equation $\partial_{\tau}{\rho}=\gamma\hat{a}\rho\hat{a}^{\dag}-{\gamma }(\hat{a}^{\dag}\hat{a}\rho+\rho\hat{a}^{\dag}\hat{a})/2$. For example, $T=1$ indicates no channel decoherence while $T=0$ does full decoherence (or full reflection). After the BS in the middle under photon losses, the total state is given by \begin{eqnarray} 
&&\ket{\Psi_{DV}^{loss}}_{ABEb\,CDEd} \nonumber \\
&&=BS^{1/2}_{B,D}BS^{T}_{B,Eb}BS^{T}_{D,Ed} \ket{\phi^{+}_{01}}_{AB} \ket{0}_{Eb}\ket{\phi^{+}_{01}}_{CD} \ket{0}_{Ed},~~~~~~
\label{Lossy_Psi_Ent01}  
\end{eqnarray}
for the presence of photon losses in modes $Eb$ and $Ed$.
The lossy environment of $\ket{\phi^{+}_{01}}_{AB}$ is mimicked by adding $BS^{T}_{B,Eb}$ with an extra vacuum state $\ket{0}_{Eb}$.
Then, if the two single-photon detections are performed at modes $B$ and $D$ on $\ket{\Psi_{DV}^{loss}}_{ABEb\,CDEd}$, the final state is given by
\begin{eqnarray}
\rho^{DV}_{AC} && = \sum_{i} p^{DV}_{i} ~\rho^{DV, i}_{AC}, \\
\rho^{DV, i}_{AC} && \propto tr_{BDEb\,Ed}\big[\hat{P}_{i} \ket{\Psi_{DV}^{loss}}_{}\bra{\Psi_{DV}^{loss}}\hat{P}_{i}\big], \label{rho_general01}
\end{eqnarray}
where $i=01,10$ and $\hat{P}_{i}s$ are from Eqs.~(\ref{proj_01_01}) and (\ref{proj_01_02}). The success probability is given by $p^{DV}_i=tr[\rho^{DV, i}_{AC}]$ for each outcome state. The total success probability  $p_{DV}$ and its entanglement negativity $E_{DV}$ on $\rho^{DV}_{AC}$ are 
\begin{eqnarray}
p_{DV}  &=& \sum_i { p^{DV}_i} = {T\over 2} (2-T), \\
E_{DV} &=&  { \sqrt{1+(1-T)^2} - (1-T)   \over 2-T},
\end{eqnarray}
which are plotted in Fig.~\ref{fig:03}.
It is interesting to note that the entanglement approaches $E_{DV} \approx (\sqrt{2}-1)/2$  for $T\approx 0$ because of $\rho^{DV}_{AC} \approx \left(\ket{00}_{AC}\bra{00} + \ket{\phi^-}_{AC}\bra{\phi^-} \right)/2$ (see Fig.~\ref{fig:03}(b)).

\subsection{Hybrid entanglement swapping with single-photon detections}
\label{Sec3-B}
Let us reuse the measurement setup of the DV BSM for a hybrid case since the parity measurement for small intensity photons is approximately equal to detecting either a vacuum or a single photon.
The initial state is given by $\ket{\Psi_{HE}}_{ABCD}=\ket{\psi_{HE}}_{AB}\ket{\psi_{HE}}_{CD}$ and the states in mode $B$ and $D$ are traveling coherent states (Fig.~\ref{scheme}(a)). Without photon losses, the total state after $BS^{1/2}$ is given by  
\begin{eqnarray}
BS^{1/2}_{B,D}\ket{\Psi_{HE}}_{ABCD}\, 
&&\propto \sum_{s=\pm} \Big[ \ket{\psi^{s}}_{AC} \ket{0}_B \ket{CS^{s}}_D \nonumber \\
&&~~~~~~+ \ket{\phi^{s}}_{AC}  \ket{CS^{s}}_B \ket{0}_D \Big],~~~~~
\label{Psi_tot_BS}
\end{eqnarray}
where $s=\pm$ and $\ket{CS^\pm} = N^{\pm}_{\sqrt{2}\alpha} \left( \ket{\sqrt{2}\alpha} \pm \ket{-\sqrt{2}\alpha} \right)$ with normalization factor $N^{\pm}_{\sqrt{2}\alpha}$.
\begin{figure}[t]
\includegraphics[width=8cm]{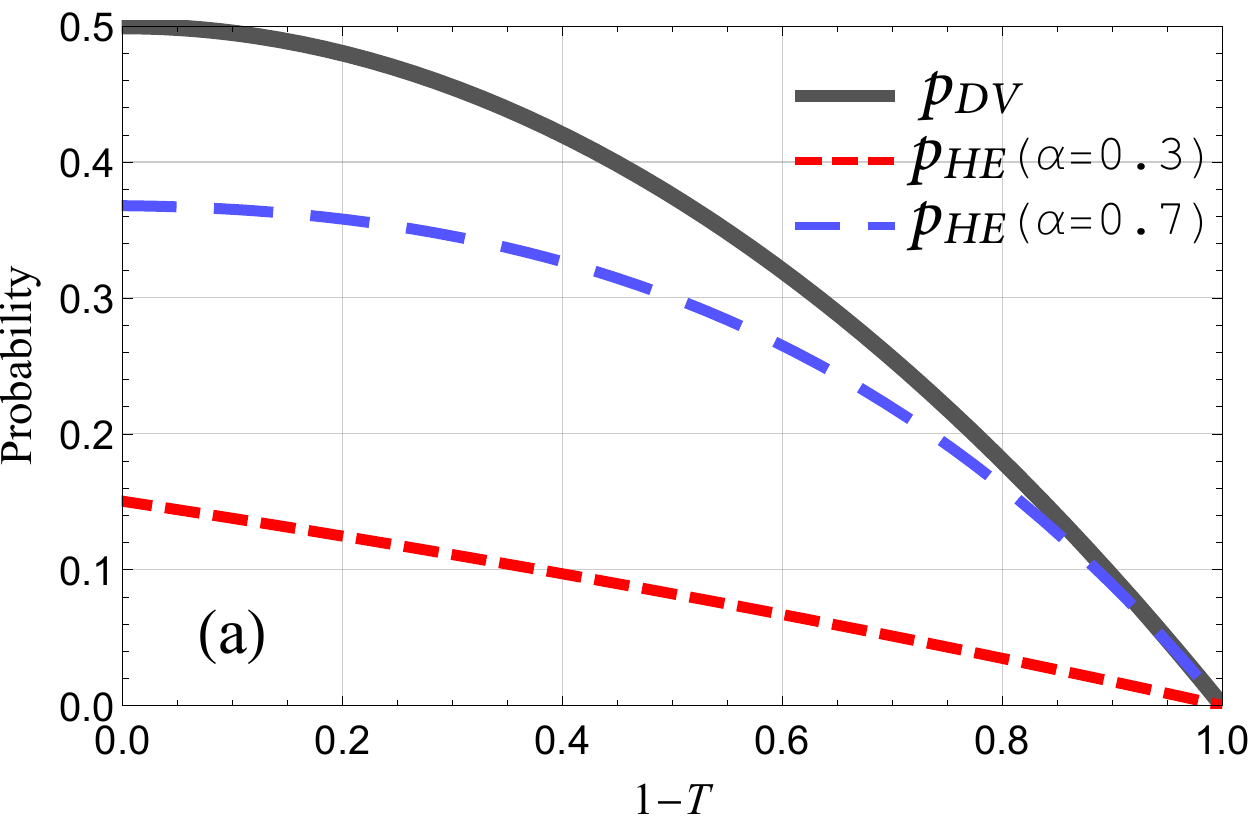}
\includegraphics[width=8cm]{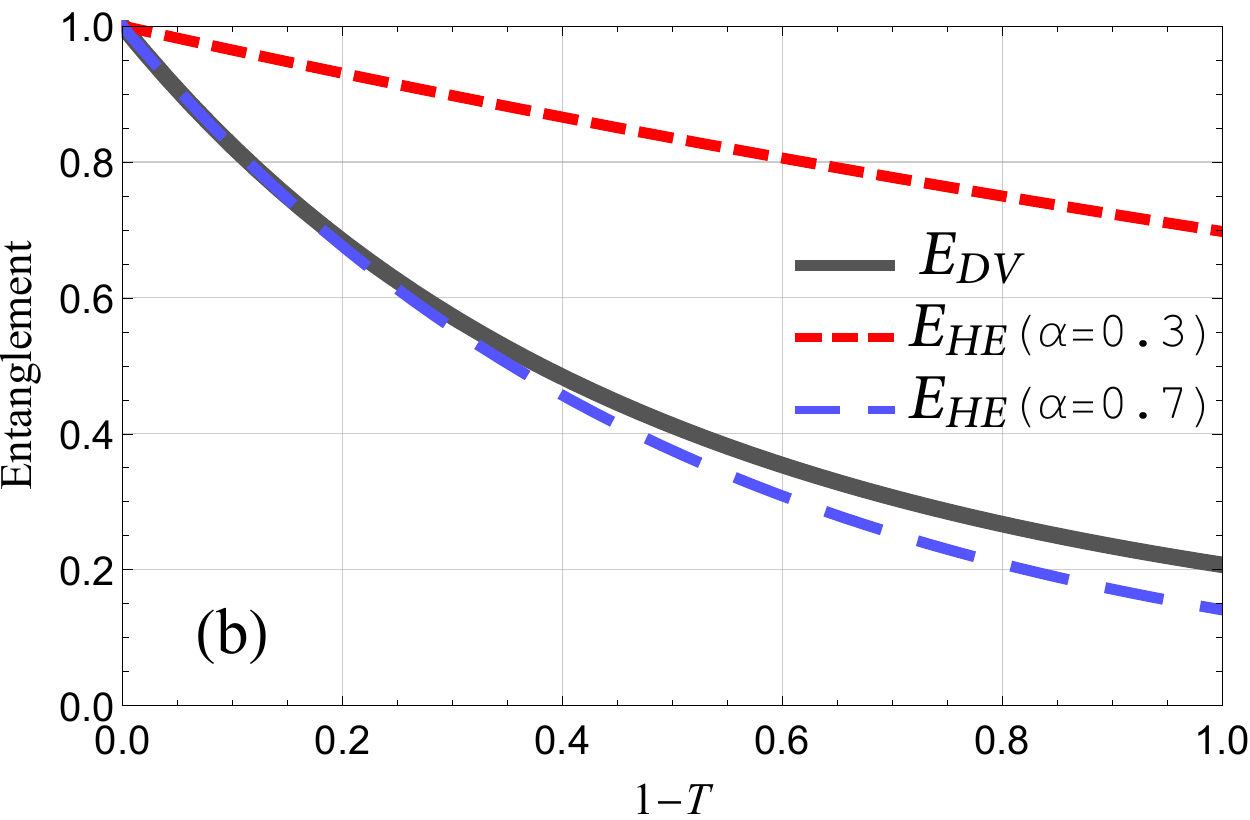}
%
\caption{(Color online) (a) Success probability and (b) entanglement negativity of the hybrid entanglement swapping as a function of the channel transmission rate $1-T$ compared with the DV entanglement swapping using VSP states. For $\alpha=0.7$, entanglement negativity of the hybrid case $E_{HE}$ overlaps with that of the DV case $E_{DV}$ for small losses $(0\leq1-T\leq0.2)$ and the success probability $P_{HE}$ is smaller than $P_{DV}$. However, for the case of small amplitudes (such as $\alpha=0.3$ in panel (b)), the amount of entanglement for the hybrid scheme is much higher than that of the DV case although it is the opposite with the success probability. } 
\label{fig:03}
\end{figure}

It shows that we achieve the unit success probability of entanglement swapping with the maximum entanglement for not too small $\alpha$ if the perfect parity measurements are performed, which distinguish among a vacuum $\ket{0}$, $\ket{CS^{+}}$, and $\ket{CS^{-}}$ in modes $B$ and $D$. However, the ideal parity measurements lose their distinguishability for small $\alpha$ due to the overlap between $\ket{0}$ and  $\ket{CS^{+}}$. Note that $\ket{CS^{-}}$ reaches to a single-photon state $\ket{1}$ for small $\alpha$ and the conclusive outcomes of the BSM with small $\alpha$ can be used for hybrid entanglement swapping {\it only if} $\ket{1} \approx \ket{CS^{-}} $ is measured in one mode and $\ket{0}$ is measured in the other mode. Thus, we examine ideal single-photon measurements for the BSM of hybrid entanglement swapping when $\alpha$ is small even in the presence of losses. Based on a similar idea of mimicking lossy channels, the final state is given by
\begin{eqnarray}
&&\rho^{HE, i}_{AC} \propto  tr_{BDEb\,Ed}\big[\hat{P}_{i}\ket{\Psi_{HE}^{loss}}_{}\bra{\Psi_{HE}^{loss}}\hat{P}_{i} \big], \label{rho01}
\end{eqnarray}
where $i=01, 10$ and
\begin{eqnarray}
\ket{\Psi_{HE}^{loss}} = BS^{1/2}_{B,D} BS^{T}_{B,Eb}BS^{T}_{D,Ed} \ket{\Psi_{HE}}_{ABCD} \ket{0}_{Eb} \ket{0}_{Ed}. \nonumber \\
\end{eqnarray}

In Fig.~\ref{fig:03}, the total success probability and entanglement for hybrid entanglement swapping in the lossy media are depicted by
\begin{eqnarray}
&&p_{HE} = 2T|\alpha|^2e^{-2T|\alpha|^2}\\
&&E_{HE} = e^{-4 (1-T) |\alpha|^2}. 
\label{suc_he}
\end{eqnarray}
The figures show the entanglement of outcome states from $E_{DV}$ has rapidly dropped with respect to $1-T$ while $E_{HE}$ slowly decreases in the lossy channel for $\alpha=0.3$. 
The success probability is relatively higher with $\alpha = 0.7$ as shown in Fig.~\ref{fig:03}(a) but the advantage of entanglement  disappears in this case as shown in Fig.~\ref{fig:03}(b).

\section{Practical scheme for hybrid entanglement swapping}
\label{Sec4}
We here consider a practical implementation of the hybrid entanglement swapping. The single-photon detection scheme can be mimicked by several on-off detectors with BSs since the photon intensity is relatively low. However, these detection schemes still give an opportunity to have a click with two or more photons even if $\alpha$ is small and the on-off detectors are perfect.
More importantly, we cannot physically detect the {\it vacuum} itself but only assume that the vacuum projection occurs when the detector is in silence. Thus, instead of the no-click event, we consider a physical setup of a POVM to detect a vacuum state. In the later subsection, we investigate the imperfect POVM and homodyne detections because the detector inefficiency critically influences the performance of entanglement swapping.

\subsection{Hybrid entanglement swapping with POVM and homodyne detection}
\label{Sec4-A} 
As shown in Fig.~\ref{scheme}(b), we modify the measurement setup in mode $B$ consisting of an additional $BS^{1/2}$ and an extra coherent state $\ket{\sqrt{2}\alpha}_E$ with two on-off detectors while a homodyne detection is performed along $\hat{x}_{\pi/2}$ in mode $D$. From Eq.~(\ref{Psi_tot_BS}), the input state is only either  $\ket{CS^{+}}_B$, $\ket{CS^{-}}_B$, or $\ket{0}_B$ in mode $B$. When the total state is given by $\ket{\Psi_{HE}^{ho}}$ = $BS^{1/2}_{BE}\,\ket{\Psi_{HE}^{loss}}\ket{0}_E$, the part of the total state in mode $B$ and $E$ is given by
\begin{eqnarray}
&& BS^{1/2}_{BE} \ket{CS^{\pm}}_B \ket{\sqrt{2} \alpha}_E \propto \ket{0}_B \ket{2 \alpha}_E \pm \ket{-2 \alpha}_E \ket{0}_B,~ \nonumber \\
&& \\
&& BS^{1/2}_{BE} \ket{0}_B \ket{\sqrt{2} \alpha}_E = \ket{- \alpha}_B \ket{ \alpha}_E.
\end{eqnarray}
Thus, if two on-off detectors successfully have a {\it click} in both modes $B$ and $E$,  the original input state in mode $B$ was not $\ket{CS^{\pm}}_B$ but a vacuum state. Thus, the final state is only collapsed into $ \ket{\psi^{+}}_{AC} \ket{CS^{+}}_D + \ket{\psi^{-}}_{AC} \ket{CS^{-}}_D $.

This POVM measurement in mode $B$ is simply described by a set of measurement (unnormalized) given by
\begin{eqnarray}
&&\hat{K}_{B}=\left\{\hat{K}_1,\hat{K}_2,\hat{K}_3,\hat{K}_4\right\},\\
&&\hat{K}_1=\ket{0}\bra{0}+\lambda^2\ket{CS^{-}}\bra{CS^{-}}-\lambda(\ket{CS^{-}}\bra{0}+\ket{0}\bra{CS^{-}}),\nonumber \\
&&\hat{K}_2=\ket{0}\bra{0}+\lambda^2\ket{CS^{-}}\bra{CS^{-}}+\lambda(\ket{CS^{-}}\bra{0}+\ket{0}\bra{CS^{-}}),\nonumber \\
&&\hat{K}_3=\ket{0}\bra{0}+\lambda^2\ket{CS^{+}}\bra{CS^{+}}+\lambda(\ket{CS^{+}}\bra{0}+\ket{0}\bra{CS^{+}}),\nonumber \\
&&\hat{K}_4=\ket{0}\bra{0}+\lambda^2\ket{CS^{+}}\bra{CS^{+}}-\lambda(\ket{CS^{+}}\bra{0}+\ket{0}\bra{CS^{+}}), \nonumber \\
&&
\end{eqnarray}
where $\lambda={\langle CS^{+} |0\rangle}=2e^{-\alpha^2}$ and $\hat{K}_4$ indicates the successful detection of $\ket{0}_B$.
When the POVM is successful, the homodyne detection along $\hat{x}_{\pi/2}$ in mode $D$ provides the final state given by
\begin{eqnarray} 
&& \ket{ \psi^{ho}}_{AC} = {1\over \sqrt{2}} (\ket{00}_{AC} +  e^{ 4i \alpha x_{\pi \over 2} } \ket{11}_{AC}).
\label{ho-result1}
\end{eqnarray}
We assume that the relative phase $e^{ 4i \alpha x_{\pi \over 2} } $ can be classically fixed by a feed-forward process given by the value $ \alpha$ and the result of homodyne measurement $x_{\pi \over 2}$. 

In the lossy case with homodyne detection, we should replace the extra coherent state to $\ket{\sqrt{2T}\alpha}$ in mode $E$ that can be determined by the channel loss rate. Then the final state is
\begin{eqnarray}
\rho^{ho}_{AC} \propto  tr_{BDE\,Eb\,Ed}\left[\hat{P}_{D}^{\pi \over 2}\hat{\Pi}_{B} \hat{\Pi}_{E} \ket{\Psi^{ho}_{HE}}_{}\bra{\Psi^{ho}_{HE}}\hat{\Pi}_{E}\hat{\Pi}_{B}\hat{P}_{D}^{\pi \over 2} \right]. \label{rho01} \nonumber \\
\end{eqnarray}
where $\hat{\Pi}=\openone - \ket{0}\bra{0}$. This projection measurement corresponds to the POVM $\hat{K}_4$, the event that both detectors click, all others are rejected by the postselection.
The corresponding success probability and entanglement negativity are
\begin{eqnarray}
&&p_{HE}^{ho} \ ={1 \over 2}(1-e^{-T|\alpha|^2})^2\label{suc_ho}\\
&&E_{HE}^{ho} \ = E_{HE} \label{neg_ho}.
\end{eqnarray}
Note that the value of entanglement with homodyne detection is equal to that with two ideal single-photon detectors due to the equivalence of two BSM setups, and we therefore confirm a practical hybrid BSM model for the ideal case with two single-photon detectors. 

\begin{figure}[t]
\includegraphics[width=8cm]{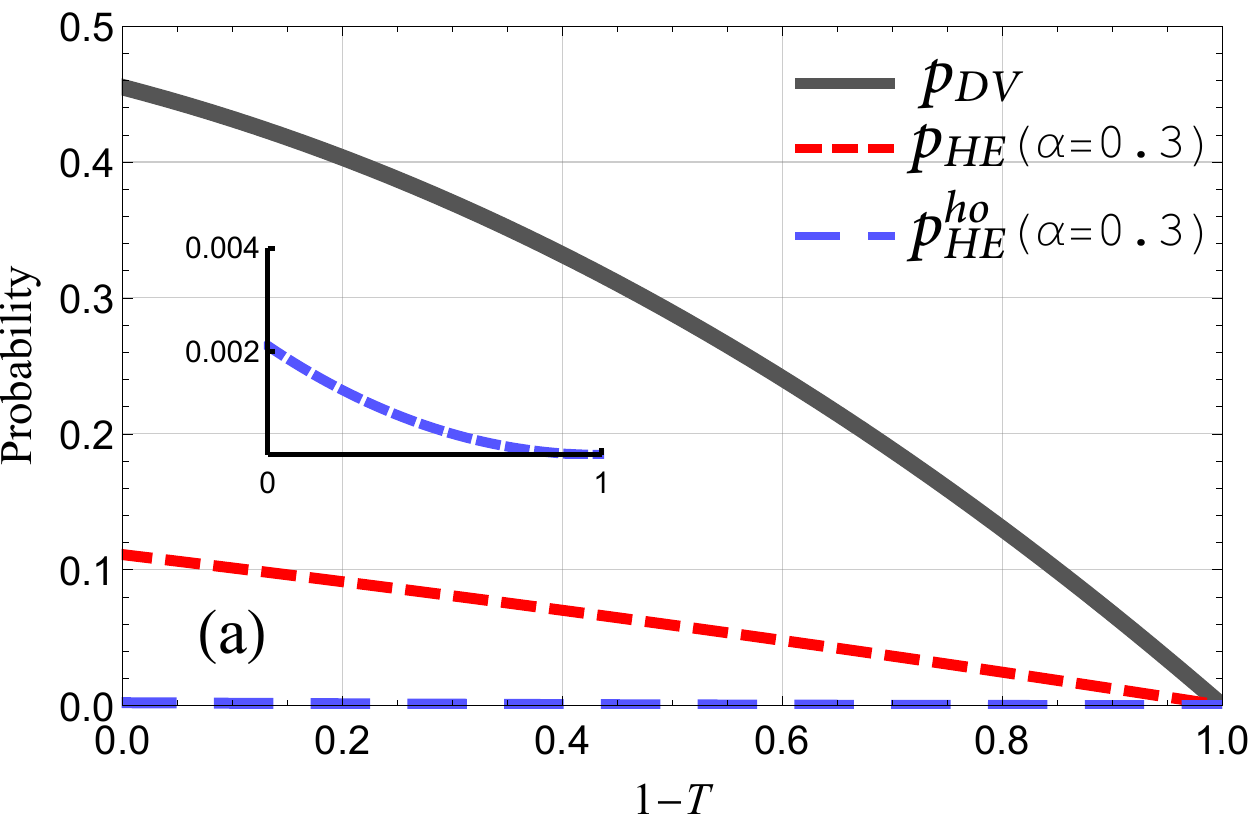}
\includegraphics[width=8cm]{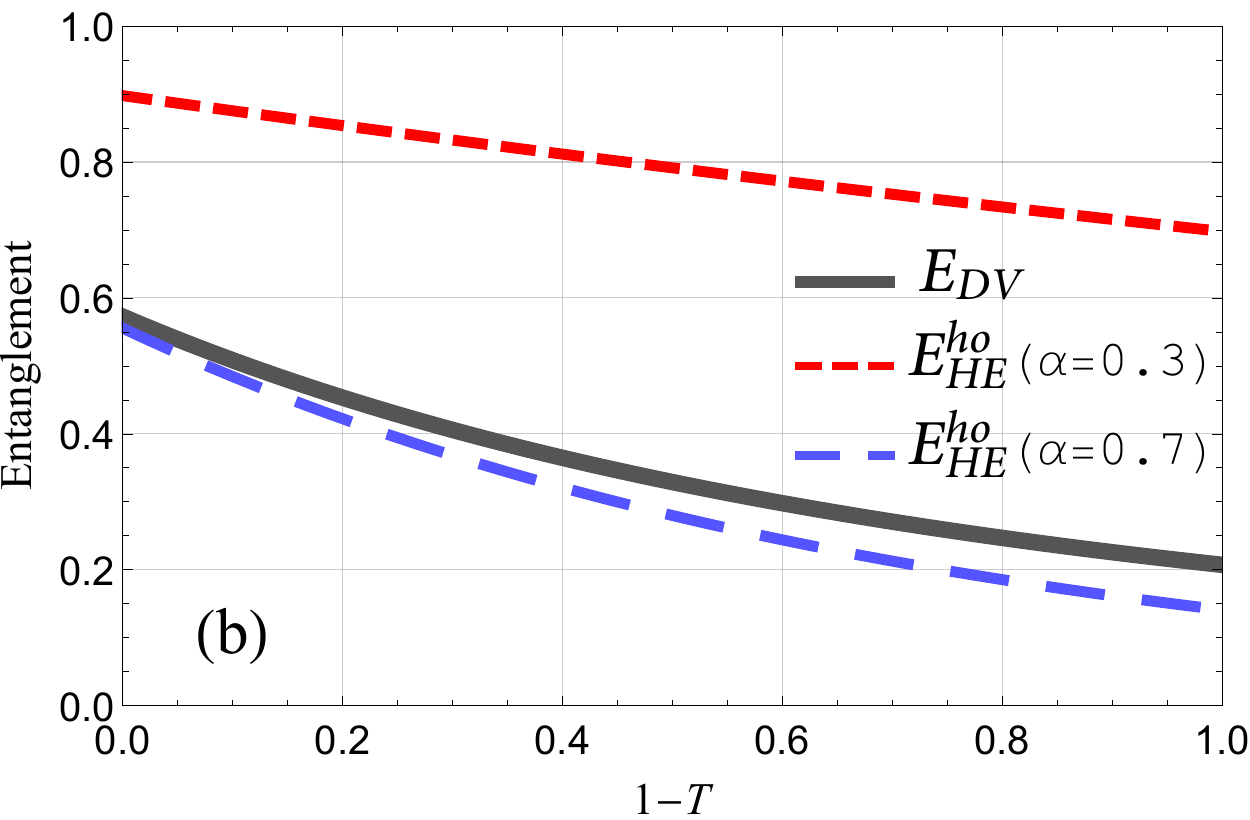}
\caption{(Color online) (a) Success probability and (b) entanglement negativity of the hybrid entanglement swapping using homodyne measurement as a function of the transmission rate $1-T$ when detection efficiency is 70\% ($T'=0.7$). In this case of imperfect detection efficiency,  entanglement of the hybrid scheme is much higher than that of the DV case for small amplitudes such as $\alpha=0.3$ in panel (b). As mentioned in the main text, the quantities of entanglement for the ideal single-photon detection and the homodyne detection are the same ($E_{HE}^{ho}=E_{HE}$).} 
\label{fig:04}
\end{figure}

\subsection{Imperfect detection}
\label{Sec4-B}
Finally, we consider the detector inefficiency by putting another $BS^{T^{\prime}}$ right before the ideal detector. It is found that the modified success probabilities and entanglement negativities are replaced by $T T^\prime$ instead of $T$ (i.e., $T^{\prime}=1$ means a perfect detection) and the results are given by
\begin{eqnarray}
&&p^{\prime}_{DV}=TT^{\prime}(2-TT^{\prime})/2, \\
&&p^{\prime}_{HE} = 2TT^{\prime}|\alpha|^2e^{-2TT^{\prime}|\alpha|^2},\label{suc_he_ineff}\\
&&p_{HE}^{\prime ho}={1 \over 2}(1-e^{-TT^{\prime}|\alpha|^2})^2, \\
&&E^{\prime}_{DV}=\sqrt{\big({1-TT^{\prime} \over2-TT^{\prime}}\big)^2+\big({1 \over2-TT^{\prime}}\big)^2} - {1-TT^{\prime} \over2-TT^{\prime}}, \\
&&E^{\prime}_{HE} = E_{HE}^{\prime ho}=e^{-4 (1-TT^{\prime}) |\alpha|^2}. \label{neg_he_ineff}
\end{eqnarray}

As shown in Fig.~\ref{fig:04}, $p_{HE}^{ho}< p_{HE} < p_{DV}$, however, the sacrifice of the success probability can be rewarded in entanglement even in the presence of both photon losses and imperfect detections. In Fig.~\ref{fig:04}(a), the inefficiency of all the detectors is fixed by $T' = 0.7$~\cite{homo1,homo2} and the success probability of  $p_{DV} < 0.5$. Fig.~\ref{fig:04}(b) shows that the advantage of entanglement quality appears with small $\alpha$ therefore it is more suitable for a practical hybrid entanglement swapping scheme.

\section{Remarks}
\label{Sec5}
We have investigated a scheme for entanglement swapping using hybrid photonic states to obtain a VSP entangled state shared by two distant parties. 
When comparing with the scheme using only VSP entangled states, our scheme shows an advantage of sharing a larger amount of entanglement with detection inefficiencies in a lossy environment. We use entanglement negativity to quantify  remaining entanglement after the entanglement swapping, which is still significantly high ($\sim 0.8$) even when the detection efficiency is 70$\%$ and the photon loss rate is 50$\%$ for $\alpha=0.3$. This value outperforms the case of DV entanglement swapping in which the entanglement is below $0.4$ under the same loss rate and detection efficiency. Moreover, our scheme gives more entanglement when $\alpha$ is small, and this is suitable for  practical implementations~\cite{Jeong2014} although there is a trade-off between the amount of entanglement and the success probability.    
This advantage of the hybrid scheme can be useful for the practical quantum key distribution (QKD) in a sense that the loss-resilient entanglement swapping protocol plays a role of the relay in QKD ~\cite{StefanoPRL, Stefano1}. A related future work may focus on the confirmation of security in QKD using this hybrid entanglement swapping scheme.

~

~

\section{Acknowledgement}
This work was supported by the National Research Foundation of Korea (NRF) grant funded by the Korea government (MSIP) (Grant No. 2010-0018295), the KIST Institutional Program (Project No. 2E26680-16-P025), and the ICT R$\&$D program of MSIP/IITP (10043464). T.P.S. acknowledges support from EPSRC (EP/M013472/1). J.J. thanks Stefano Pirandola for useful discussions.

\end{document}